\documentclass[aps,prl,twocolumn,groupedaddress]{revtex4-1}

\usepackage{amsmath}

\usepackage{graphicx}

\DeclareGraphicsExtensions{.pdf,.eps,.png,.jpg,.mps}

\usepackage{epstopdf}

\usepackage{threeparttable}

\usepackage{color}

\begin{document}

\title{Direct observation of intravalley spin relaxation in single-layer WS$_2$}

\author{Z. Wang$^1$, A. Molina-Sanchez$^2$, P. Altmann$^1$, D. Sangalli$^3$, D. De Fazio$^4$, G. Soavi$^4$, U. Sassi$^4$, F. Bottegoni$^1$, F. Ciccacci$^1$, M. Finazzi$^1$, L. Wirtz$^5$, A.C. Ferrari$^4$, A. Marini$^3$, G. Cerullo$^{1,6}$, S. Dal Conte$^1$}

\affiliation{$^1$ Dipartimento di Fisica, Politecnico di Milano, Piazza L. da Vinci 32, I-20133 Milano, Italy}

\affiliation{$^2$ Institute of Materials Science (ICMUV), University of Valencia, Catedr\'{a}tico Beltr\'{a}n 2, E-46980, Valencia, Spain}

\affiliation{$^3$ CNR-ISM, Division of Ultrafast Process in Materials (FLASHit), Area della Ricerca di Roma 1, Monterotondo Scalo, Italy}

\affiliation{$^4$ Cambridge Graphene Centre, University of Cambridge, 9 JJ Thomson Avenue, Cambridge CB3 0FA, UK}

\affiliation{$^5$ Universit\'e du Luxembourg, 162 A, avenue de la Faencerie, L-1511 Luxembourg}

\affiliation{$^6$ IFN-CNR, Piazza L. da Vinci 32, I-20133 Milano, Italy}

\begin{abstract}

In monolayer Transition Metal Dichalcogenides (TMDs) the valence and conduction bands are spin split because of the strong spin-orbit interaction. In tungsten-based TMDs the spin-ordering of the conduction band is such that the so-called dark exciton, consisting of an electron and a hole with opposite spin orientation, has lower energy than the A exciton. A possible mechanism leading to the transition from bright to dark excitons involves the scattering of the electrons from the upper to the lower conduction band state in K. Here we exploit the valley selective optical selection rules and use two-color helicity-resolved pump-probe spectroscopy to directly measure the intravalley spin-flip relaxation dynamics of electrons in the conduction band of single-layer WS$_2$. This process occurs on a sub-ps time scale and it is significantly dependent on the temperature, indicative of a phonon-assisted relaxation. These experimental results are supported by time-dependent ab-initio calculations which show that the intra-valley spin-flip scattering occurs on significantly longer time scales only exactly at the K point. In a realistic situation the occupation of states away from the minimum of the conduction band leads to a dramatic reduction of the scattering time.

\end{abstract}

\maketitle

\begin{figure}

\includegraphics[width=60mm]{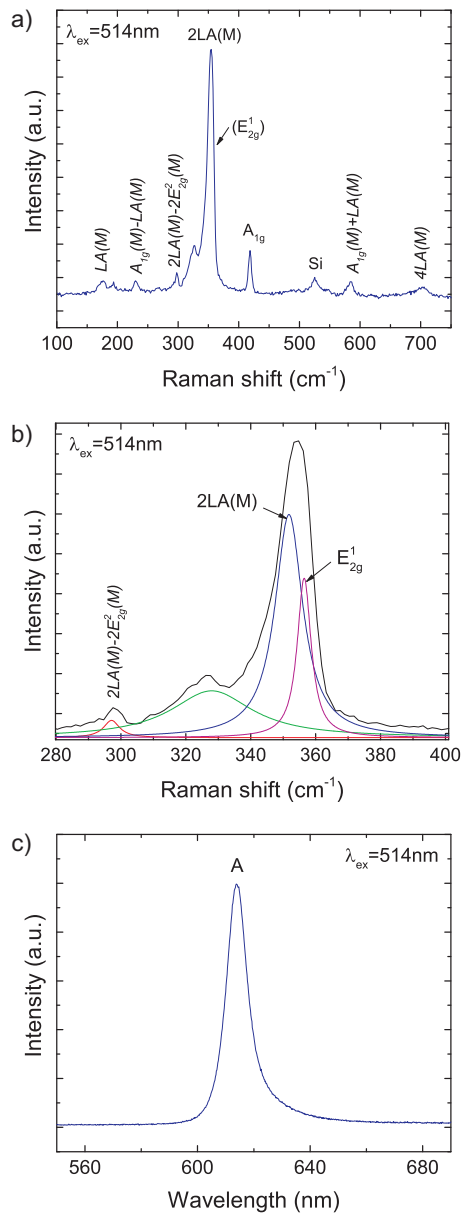}

\caption{(a) Raman spectrum of 1L-WS$_2$ measured with the pump at 514nm (b) Close up of the Raman spectrum showing the multi-peak fit (c) PL spectrum of 1L-WS$_2$ pumped at 514nm at room temperature.}

\label{FigSP1}

\end{figure}

Transition metal dichalcogenides (TMDs) are promising for opto-electronics\cite{Wang2012,Koppens2014,Mak2016,Ferrari2015}, valleytronics\cite{Yuan2014,Yao2008,Aivazian2015,Ye2016,Schaibley2016} and quantum information processing\cite{Srivastava2015}. Monolayers (1L) of TMDs are direct bandgap semiconductors\cite{Mak2010} whose optical properties are dominated by excitons with binding energies of up to several hundred meVs\cite{Chernikov2014,Ugeda2014,Zhu2015,Komsa2012,Qiu2013,Molina-Sanchez2015,Molina-Sanchez2016}. The valence/conduction (VB/CB) band extrema lie at the non-equivalent K and K' points on the edge of the Brillouin zone\cite{Zhang2014,Splendiani2010}. Their spin-degeneracy is lifted by strong spin-orbit (SO) interaction\cite{Molina-Sanchez2013}. For the VB the energy splitting $\Delta_v$ ranges between 150 and 400meV\cite{Li2014}, while for the CB $\Delta_c$ it is one/two orders of magnitude lower ($\sim$1-30meV)\cite{Kosmider2013,Kormanyos2014}. Together with a broken inversion symmetry, this results in spin-polarized bands and valley-dependent dipole-allowed interband optical transitions, as first detected by helicity resolved photoluminescence (PL) measurements\cite{Mak2012,Cao2012,Zeng2012}. The large VB splitting gives rise to two distinct interband transitions (called A and B) strongly renormalized by excitonic effects\cite{Qiu2013}, dominating the optical response in the visible range\cite{Li2014}.\\
Recent theoretical calculations predicted that, while the spin orientation of the upper and lower CB states is antiparallel in K/K', in W-based 1L-TMDs, the upper CB state has the same spin orientation as the upper VB. A direct consequence of this spin-ordering is the formation of an intravalley (i.e. zero momentum) dark exciton with a lower energy than the bright exciton\cite{Yu2015,Zhang2015,Baranowski2017,Slobodeniuk2016,Jin2017,ZhangNN2017}. The bright to dark exciton transition reduces the PL quantum yield in W-based 1L-TMDs for decreasing temperatures, as recently shown\cite{Zhang2015}. The formation of dark excitons requires scattering processes such as intravalley relaxation. We stress that other excitonic complexes with a non-zero momentum are referred to as dark\cite{Malic2018}, but are beyond the scope of this study.\\     
The question about how excitons scatter to other CB and VB states is highly discussed within the community. Most of the experiments, performed up to now on TMDs, focused on the study of intervalley scattering processes\cite{DalConte2015,Zhu2014,Mai2014,Mai2014a}. On the contrary, the intravalley scattering process is not well understood. On the one hand Refs.~\cite{Plechinger2016,Plechinger2017} suggested that the intravalley scattering from the upper to the lower spin-split CB is one of the possible decay channels for the A exciton Kerr signal. On the other hand, based on the assumption that a spin-flip event occurs on a significantly longer time scale\cite{Song2013,Ochoa2013,Wang2014}, the intravalley scattering is neglected in recent theoretical models describing exciton scattering in terms of electron-hole exchange interaction\cite{Glazov2014}, Dexter mechanism\cite{Berghaeuser2018} (i.e. a Coulomb-induced intervalley coupling between the A and B exciton) and upconversion process\cite{Manca2017}.\\

\begin{figure}

\includegraphics[width=70mm]{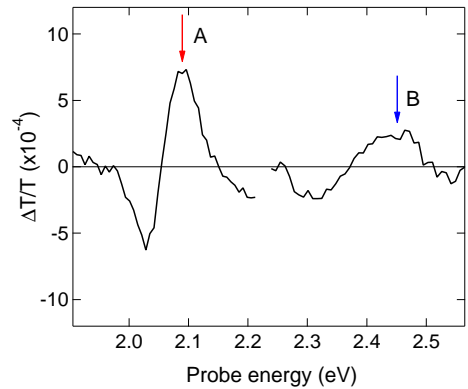}

\caption{Energy resolved $\Delta T/T$ spectra at a fixed delay time $\tau$=100ps covering the A and B exciton transitions at 77K. Both pump and probe pulses are linearly polarized. The red and blue arrows, centered around the maximum of the bleaching signal, indicate the pump and probe energies, respectively.}

\label{Fig:BroadbandAbsorption}

\end{figure}

\begin{figure*}

\centerline{\includegraphics[width=170mm]{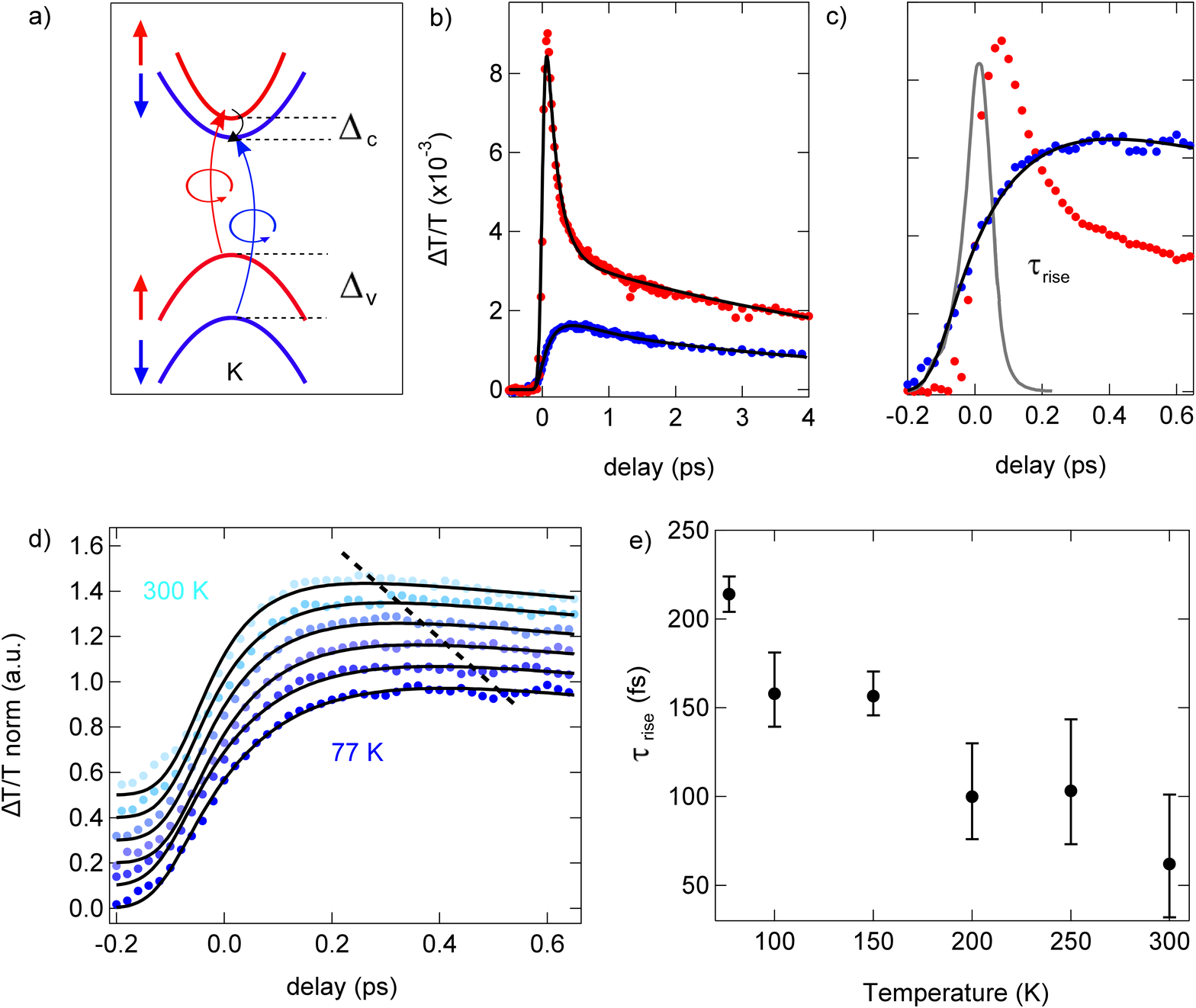}}

\caption{Intravalley relaxation dynamics in 1L-WS$_2$. (a) Spin-flip relaxation process. A excitons are injected in the K valley by a circularly polarized pump (red arrow). The probe pulse has the same helicity and it is resonant with the B excitonic transition (blue arrow). The gradual filling of the lower CB states due to intravalley scattering hinders the absorption of the probe by Pauli blocking, causing a delayed formation of the bleaching signal around the B exciton. (b) Red and blue curves are $\Delta T/T$ at 77K around the A and B transitions. (c) Close up at shorter delay time of the transient optical response. The blue trace is multiplied by 4 to highlight the different build-up dynamics. For the same pump and probe energies, the build-up dynamics is pulsewidth limited, while the non-degenerate configuration displays a delayed formation. The grey line is the cross correlation between pump and probe pulses. The black line is the fit to the data. (d)$\Delta T/T_{B,K}$ at different temperatures. The dashed line through the maximum of each trace highlights the increasing rate of intravalley scattering at higher temperatures. (e) Temperature dependence of $\tau_{rise}$. The error bars are derived from the fit of the $\Delta T/T$ traces reported in (d).}

\label{Fig:IntravalleyDynamics}

\end{figure*}

Here we present a combined experimental and theoretical study of the intravalley scattering in 1L-WS$_2$. Exploiting two-color helicity-resolved pump-probe spectroscopy, we find an almost instantaneous build-up of the occupation of both CB states in the K' valley after excitation of the A-exciton transition within the K valley. The build-up of the occupation of the lower CB state within the same valley, instead, is clearly delayed; however, it still displays a surprisingly short time constant of $\sim$200fs at 77K. We find that this process becomes faster at higher temperature indicating a phonon-mediated spin-flip scattering.\\
This interpretation is confirmed by time-dependent ab-initio calculations\cite{Marini2013} where the non-equilibrium many body perturbation theory (MBPT) is combined with density functional theory (DFT) \cite{Onida2002} to describe electron-electron and electron-phonon scattering. We show that the previously assumed long spin-flip scattering processes occur only at the exact minimum of the CB. All the excitation processes involving electronic states nearby the K point, induced for instance by a background doping or by the finite energetic bandwidth of the pump laser, drastically decrease the scattering time. These results are of great importance for spintronics devices based on 1L-TMDs\cite{Schaibley2016,Gmitra2015}.

1L-WS$_2$ flakes are produced by micromechanical cleavage from bulk WS$_2$ crystals (HQ Graphene, The Netherlands). These are exfoliated with Nitto Denko tape, then further exfoliated on a polydimethylsiloxane (PDMS) stamp for inspection under an optical microscope and stamping on a 200$\mu$m-thick fused silica substrate, as described in Ref.\cite{Castellanos-Gomez2014}. In order to facilitate the flakes' location, a metallic frame is fabricated around them by laser-writer lithography, evaporation of Cr/Au and lift-off. The flakes are characterized by Raman spectroscopy and PL using a Renishaw InVia spectrometer with a 100$\times$ objective and 514nm excitation. In WS$_2$ the 2LA(M) mode at$\sim$352cm$^{-1}$, due to the longitudinal acoustic phonon at the M point, overlaps with the E$^1_{2g}$\cite{Berkdemir2013}, requiring a multiple-Lorentzian fit to resolve them, Fig.\ref{FigSP1}b. The intensity ratio of the 2LA(M) to the A$_{1g}$ mode at 419cm$^{-1}$ is I(2LA(M))/I(A$_{1g}$)$\sim$4.6, consistent with 1L-WS$_2$\cite{Berkdemir2013}. Fig.\ref{FigSP1}c shows a single PL peak appearing at$\sim$614nm ($\sim$2.02eV). This peak originates from the radiative recombination of the A exciton. No trace of light emission from the indirect gap, at energy lower than A exciton, is detected confirming the 1L structure of the sample\cite{Gutierrez2013}.

Both the equilibrium and non-equilibrium optical response of 1L-WS$_2$ are dominated by two excitonic resonances\cite{Li2014}. In order to precisely determine the position of these peaks, we perform broadband transient absorption measurements on an energy range covering the visible spectrum. The laser used in this experiment is an amplified Ti:Sapphire laser (Coherent Libra), emitting 100fs pulses at 800nm, with an average power of 4W at 2kHz repetition rate. A fraction (i.e. 300mW) of the total laser power is used for the experiment. The output beam is equally divided by a beam splitter into two parts serving as a pump and probe, respectively. The pump beam is generated by an home-built non-collinear optical parametric amplifier (NOPA) pumped by the second harmonic of the laser output, it has a narrow band (10nm) and can be continuously tuned in the entire visible range (from 1.6eV to 2.5eV). The probe beam is obtained by white light continuum generation in a 2mm-thick sapphire plate. Pump and probe pulses are collinearly focused on the sample with an achromatic doublet. The pump pulse is modulated at 1kHz by a mechanical chopper and, at the sample position, has a diameter of 12$\mu m$. After the interaction with the sample, the probe beam is dispersed by a prism and is detected by a Si CCD camera. In this experiment, the pump beam is linearly polarized and it is tuned slightly above each of the A/B exciton peaks while the white-light probe has perpendicular polarization and has a broad spectral content covering both exciton transitions.
Figure~\ref{Fig:BroadbandAbsorption} shows the $\Delta T/T$ spectrum at fixed delay ($\tau=100$~ps), which exhibits a characteristic spectral shape previously observed in other monolayer TMDs~\cite{Pogna2016,Sim2013,Vega-Mayoral2016}. In general, the transient optical response of TMDs close to the excitonic peaks, is the consequence of complex many-body processes taking place after photoexcitation. When the pump pulse induces an electronic optical transition, phase-space filling effects due to Pauli-blocking reduce the absorption close to the band-edge and such induced transparency as probed by the second pulse results in a photo bleaching signal (positive $\Delta T/T$). Meanwhile the photoinduced variation of the Coulomb screening of the electron-electron and electron-hole interactions renormalizes the quasi-particle gap and the exciton binding energy, respectively, resulting in transient redshift of the absorption edge of the order of tens of meV~\cite{Ruppert2017}. Moreover, excitation induced dephasing processes leading to a transient broadening of the absorption peak, can also take place as recently shown in WS$_2$~\cite{Ruppert2017,Sie2017}. All these aforementioned processes compete against each other and dominate the 1L-WS$_2$ transient optical response within the first picoseconds. On a longer time scale (i.e. $10$~ps), the system thermalizes with the phonon population~\cite{Ruppert2017}, and both excitonic peaks shift back to their original positions. Therefore, from the maximum of the photo bleaching signal at long delay time, it is possible to give a solid estimation of the spectral positions of the A and B excitonic resonances which are $E_A=2.09$~eV and $E_B=2.45$~eV, respectively.

We then perform transient absorption measurements with the pump tuned to $E_A$, while the probe is either resonant with $E_A$ (degenerate configuration) or $E_B$ (non-degenerate configuration) (Fig.\ref{Fig:IntravalleyDynamics}a). In this measurement, the pump and the probe beam are generated by two NOPAs and they have narrow spectral content. Both the beams are circularly polarized by a broadband quarter wave-plate (BHalle). After the interaction with the sample, the probe beam is detected by a photodiode. The readout of the photodiode is then demodulated by a lock-in amplifier (Stanford SR830) enabling the detection of differential transmission $\Delta T/T$ signals down to 10$^{-4}$. The spin ordering of the CB states in 1L-WS$_2$ is such that the pump excites electrons from the upper VB to the upper CB. The photo-generated holes are not expected to scatter into the lower VB state within the time scale of the experiment, because of the large splitting in the VB\cite{Mak2012}. 
Thus, in the degenerate configuration, the probe is sensitive to the decay of the excited electron population in the upper CB. In the non-degenerate configuration, the probe measures the electron population in the lower CB, which is the one related to the dark excitons. 
With left- (right-) circular polarized light, we discretely access the K (K') valley, because of the optical selection rules. This enables us to disentangle the intra- and intervalley scattering processes. 
To study the intravalley relaxation dynamics, both pump and probe, have the same helicity. 
Fig.\ref{Fig:IntravalleyDynamics}b shows the transient absorption signal for the degenerate (red curve, $\Delta T/T_{A,K}$) and non-degenerate (blue curve, $\Delta T/T_{B,K}$) configuration. The positive sign of both curves corresponds to a photobleaching of the A and B transitions. 
The $\Delta T/T_{A,K}$ trace exhibits a pulse-width limited build-up followed by a relaxation dynamics fitted by a double exponential curve with decay times $\tau_A^{fast}=150\pm10$fs and $\tau_A^{slow}=4.5\pm0.5$ps. $\Delta T/T_{B,K}$, instead, exhibits a single exponential decay with $\tau_B=5\pm 0.5$ps (Fig.\ref{Fig:IntravalleyDynamics}b) and reaches its maximum at a delayed time with respect to the pump excitation, as clearly seen in Fig.\ref{Fig:IntravalleyDynamics}c. The rise time $\tau_{rise}=210\pm 10$fs is estimated by fitting the build-up dynamics with the function $1-exp(-t/\tau_{rise})$ convoluted with a Gaussian instrumental response function (IRF) obtained by sum-frequency cross-correlation experiments between pump and probe pulses accounting for the temporal resolution of 100fs.\\ 
The instantaneous rise of $\Delta T/T_{A,K}$ is interpreted as a phase space filling of the final state by Pauli blocking and transient optical gap renormalization. 
The latter is caused by the transient change of the Coulomb screening and has a characteristic timescale (i.e. tens of femtoseconds\cite{Huber2001}) much faster than the temporal resolution of the setup (100fs).Therefore, the finite build-up of the photobleaching signal around the B exciton can only arise from a Pauli blocking effect, due to the scattering of the carriers from the upper to the lower CB in K.

Many scattering mechanisms contribute to the depletion of the bright state on a sub-ps timescale, since bright excitons can radiatively recombine in the same valley\cite{Zhu2014}, scatter into the opposite valley\cite{Mai2014,Mai2014a,DalConte2015,Manca2017,Berghaeuser2018} or form dark excitons with non-zero momentum\cite{Selig2018}. 
It is believed that the direct intravalley scattering, because it requires a spin-flip, occurs on times $>10ps$\cite{Ochoa2013,WangPRB2014,Ochoa2013_2,Ochoa2014,SchmidtPRL2016,Song2013}. 
Our experiment, however, suggest that this process might be much faster. We find that $\tau_A^{fast}$ (150fs) is very close to $\tau_{rise}$ (210fs). Therefore, $\tau_{rise}$ can be associated with the intravalley scattering time. Since intravalley relaxation is a transition of electrons from the upper to the lower CB, we expect the scattering time to depend on the spin splitting $\Delta_c$ which, for 1L-WS$_2$, is estimated to be $\sim$26meV\cite{Nota}.

\begin{figure}

\includegraphics[width=70mm]{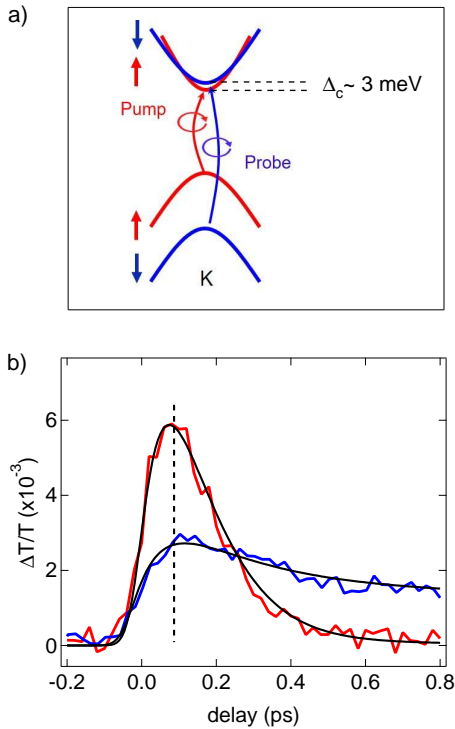}

\caption{(a)1L-MoS$_2$ band structure around K. $\Delta_C$ is the CB splitting. (b)$\Delta T/T$ around the A (red curve) and B (blue curve) excitons at 77K. The incident pump fluence is the same as that for the $\Delta T/T$ measurements on 1L-WS$_2$ in Fig.1 of the main text. The continuous lines are the fit to the data. Both traces displays the same build up time as indicated by the dashed line.}

\label{Fig:IntraMoS2}

\end{figure}

\begin{figure*}

\centerline{\includegraphics[width=170mm]{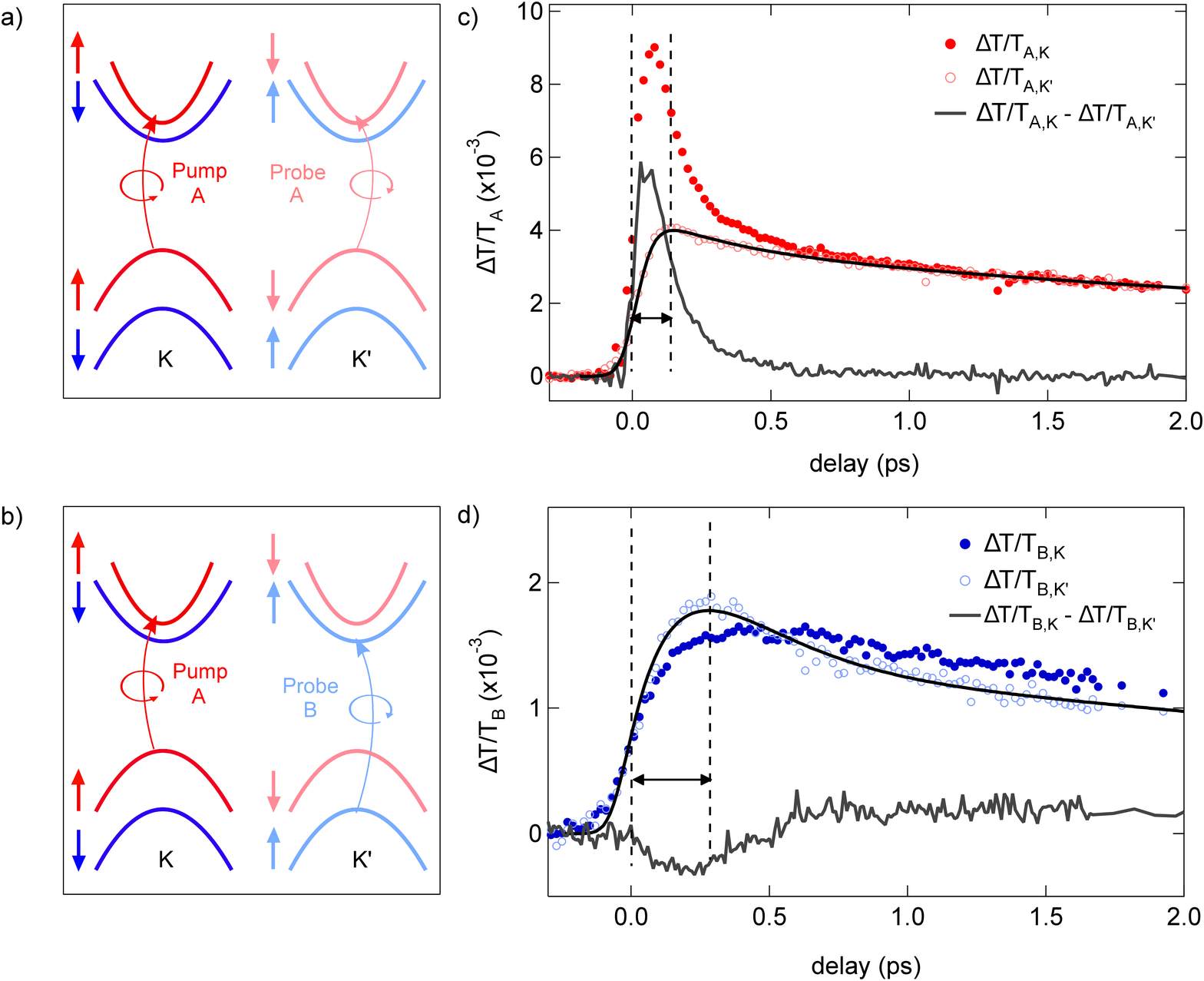}}

\caption{Intervalley relaxation dynamics in 1L-WS$_2$. (a,b) Intervalley scattering processes. The K' bands have opposite spin orientation and are depicted with lighter colors. (c,d) Temporal dynamics obtained by photoexciting A excitons in K and measuring $\Delta T/T$ in K and K' at the A and B resonances at 77K. The grey lines are the difference between the signals with same and opposite polarizations. They are directly related to the valley depolarization dynamics. The dashed lines are centered at zero delay and at the maximum of the $\Delta T/T$ in K'.}

\label{Fig:IntervalleyDynamics}

\end{figure*}

\begin{figure*}

\centerline{\includegraphics[width=170mm]{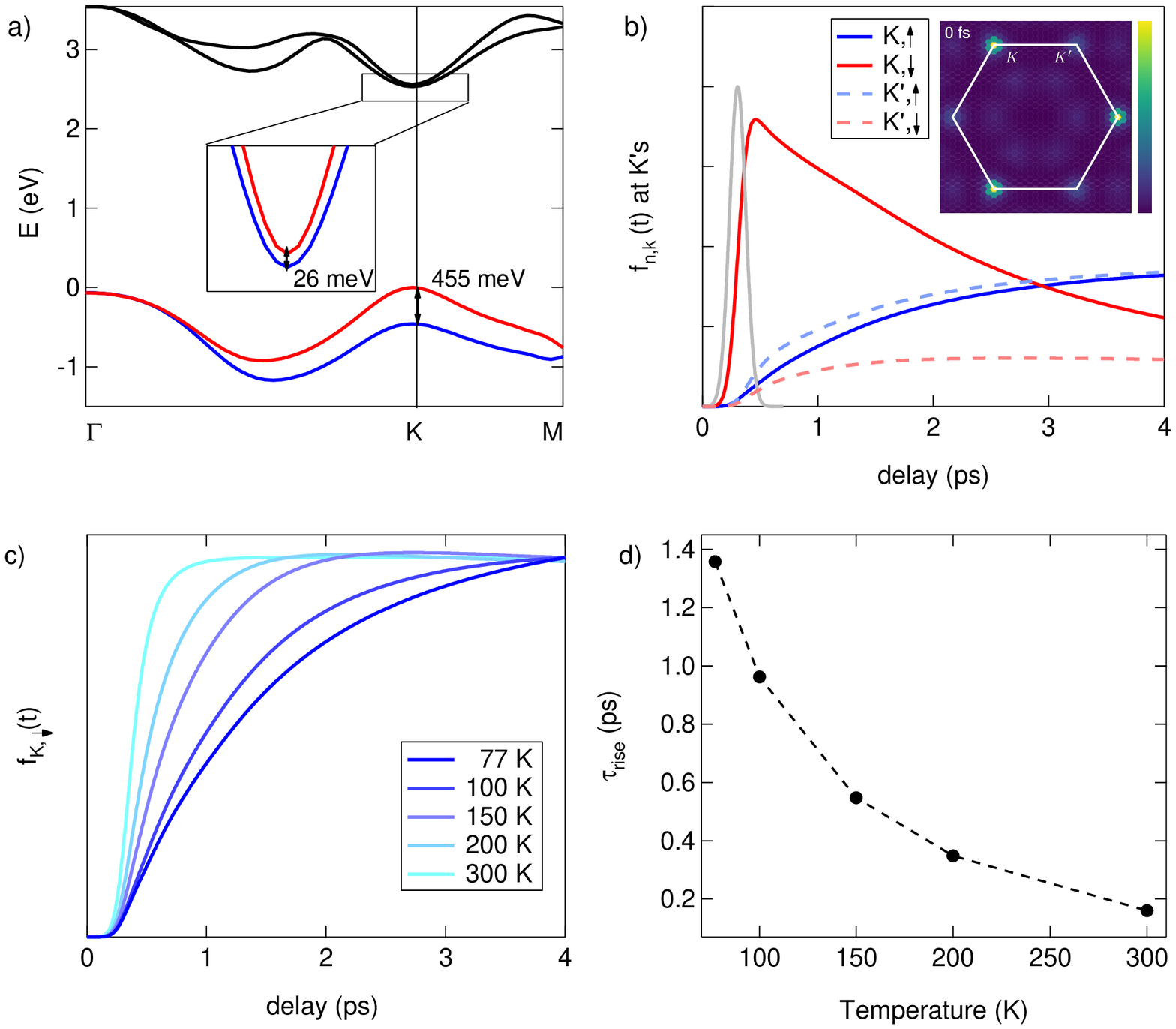}}

\caption{(a)1L-WS$_2$ band structure. The inset shows the CB near K. The red and blue colors corresponds to $S_z=\uparrow$ and $S_z=\downarrow$. (b) Time evolution of the carrier population in the (K,$\uparrow$) (K,$\downarrow$), (K',$\uparrow$) and (K',$\downarrow$) states. The inset depicts the carrier distribution at zero pump-probe delay projected on the K,K' momentum plane. At $\tau$=0 the K valleys concentrate most of the electron population. (c) Time evolution of the K,$\downarrow$ population as a function of the temperature. (d) Corresponding rise time.}

\label{Fig:theory}

\end{figure*}

This interpretation is substantiated by comparing these findings to the case of MoS$_2$. Here the spin ordering of the CB state in K is reversed compared to 1L-WS$_2$ implying that bright exciton has lower energy than dark exciton\cite{Kosmider2013}. Moreover ref.\cite{Kosmider2013} shows that, in 1L-MoS$_2$, chalcogen and metal atoms contributions to CB splitting have opposite signs and almost cancel each other, resulting in a CB splitting extremely small ($\sim$3meV\cite{Kosmider2013}) compared to the other 1L-TMDs. Fig.\ref{Fig:IntraMoS2} reports the transient optical response measured around the A ($E_A=1.85eV$) and B ($E_B=2.05eV$) excitons of 1L-MoS$_2$ (red and blue curves respectively) after that spin polarized carriers are photoinjected in the K valley by a 100fs pump pulse in resonance with the A exciton. Unlike 1L-WS$_2$, both $\Delta T/T$ traces have the same pulsewidth-limited build up dynamics. This shows that the intravalley spin flip scattering process is more efficient for 1L-MoS$_2$ than 1L-WS$_2$. This result confirms the trend observed in the theoretical simulations by ref.\cite{Song2013} where it has been predicted that, despite the lower SO splitting, Mo-based 1L-TMDs have faster spin relaxation time than W-based 1L-TMDs, and further supports the interpretation that $\tau_{rise}$ is a direct measurement of the intravalley scattering time.

\begin{figure}

\includegraphics[width=90mm]{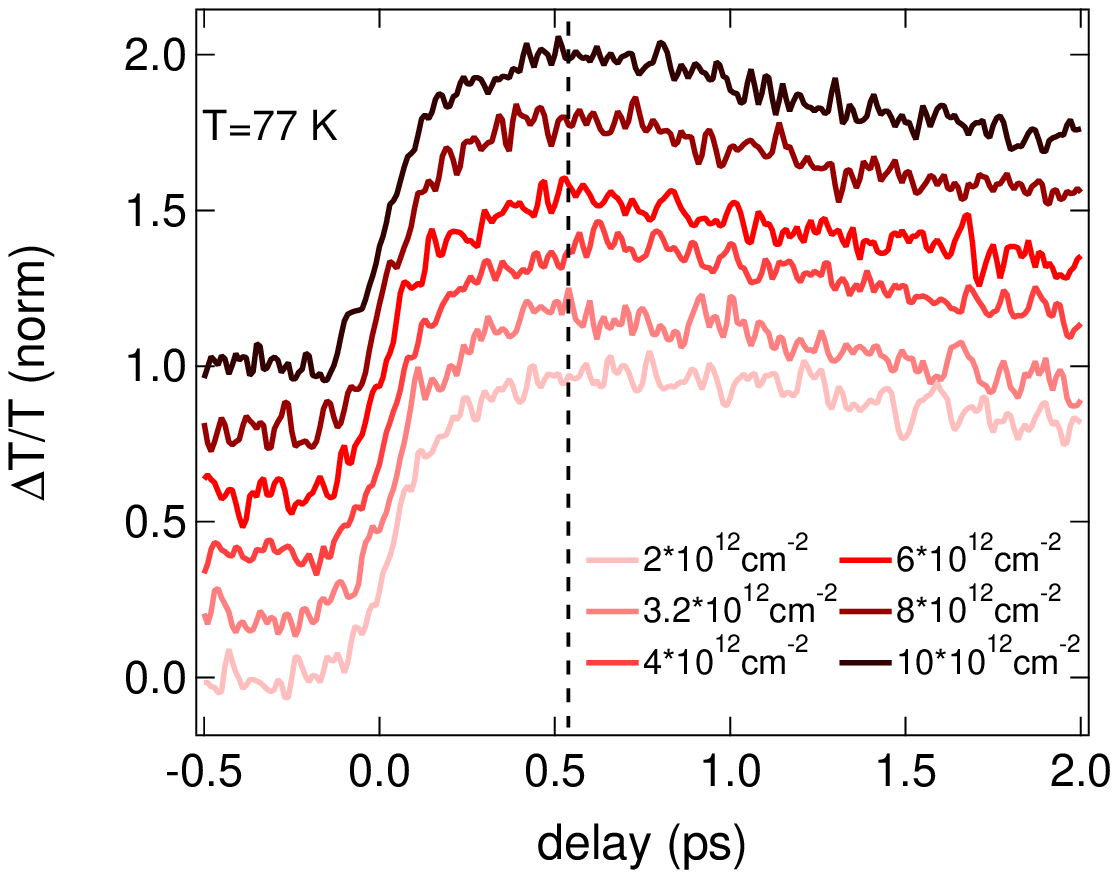}

\caption{Normalized $\Delta T/T$ around the B exciton peak in 1L-WS$_2$ at 77K for different photoinduced carrier densities.}

\label{Fig:Fluencedep}

\end{figure}

To gain a deeper insight into the underlying scattering mechanism, we study the temperature dependence of $\Delta T/T_{B,K}$ (Fig.\ref{Fig:IntravalleyDynamics}d). $\tau_{rise}$ significantly decreases at higher temperature (Fig.\ref{Fig:IntravalleyDynamics}e). This suggests that the intravalley spin relaxation between upper and lower CB is a phonon-assisted process. 
Among all the first order phonon scattering processes, only phonons with a momentum close to zero (around $\Gamma$), can play an active role. The $\Gamma$ in-plane, Raman-active, optical phonons with symmetry $\Gamma_{5}$ at $\sim$285cm$^{-1}$ are good candidates for mediating such a scattering process within the same valley because their energy ($\sim$35meV) is very close to the spin splitting in the CB\cite{Molina-Sanchez2011}. 
Long wavelength flexural phonons have been considered theoretically, but the associated relaxation dynamics is orders of magnitude longer than in our measurements\cite{Song2013,Ochoa2013_2}.\\

In addition to intravalley relaxation, scattering into the opposite valley is expected to be a major channel for the depletion of the A excitons in K\cite{Mai2014,Mai2014a,DalConte2015,Manca2017,Berghaeuser2018}. This is addressed, in our experiment, by switching the probe to the opposite helicity. 
We measure the transient absorption in the K' valley around the A ($\Delta T/T_{A,K'}$) and B ($\Delta T/T_{B,K'}$) transitions, as sketched in Fig.\ref{Fig:IntervalleyDynamics}a,b. 
In Fig.\ref{Fig:IntervalleyDynamics}c, we compare $\Delta T/T_{A,K'}$ (pink) to the previously discussed instantaneous build-up in $\Delta T/T_{A,K}$. A $\tau_{rise}=30\pm 10$fs can be estimated, despite the finite (i.e. 100fs) temporal resolution. This fast timescale is consistent with the high-temporal resolution time-resolved photoemission experiments of Ref.\citenum{Bertoni2016}, confirming the strong coupling between K and K' valleys. 

The valley polarization is defined as the imbalance of the exciton population between K and K', i.e. ($\Delta T/T_{A,K}$-$\Delta T/T_{A,K'}$). The A polarization (grey line in Fig.\ref{Fig:IntervalleyDynamics}c) drops by a factor 10 within the first $\sim$500fs, followed by a slower decay within a few picoseconds, in agreement with previous measurements on 1L-MoS$_2$\cite{DalConte2015} and 1L-WS$_2$\cite{Berghaeuser2018}. 
This fast depolarization has been previously explained by electron-hole exchange interaction\cite{Yu2014a,Yu2014,Manca2017} and Coulomb induced\cite{Schmidt2016} or phonon mediated\cite{Carvalho2017} intervalley scattering.
We stress that the robustness of the valley polarization is also limited by intravalley scattering, which manifests on the subpicosecond timescale. 

Fig.\ref{Fig:IntervalleyDynamics}d compares the measurements of the B transition in the two valleys. The rise time of the signal in the unpumped valley is found to be $90\pm 20$fs. Interestingly, despite the large momentum mismatch, both $\Delta T/T_{A,K'}$ and $\Delta T/T_{B,K'}$, exhibit a faster build-up than $\Delta T/T_{B,K}$. 
The valley polarization of the B exciton (grey curve) shows a peculiar sign change, which has been previously observed\cite{Berghaeuser2018} and can be explained by the slightly different timescales of inter- and intravalley scattering.

We test our hypothesis on the fast intravalley relaxation process by time-dependent ab-initio calculations using a first-principles implementation of non-equilibrium MBTP\cite{Marini2013,Sangalli2015a,Attaccalite2011,Perfetto2015,MolinaSanchez2017}. 
The band structure and the static absorption of 1L-WS$_2$ are calculated using equilibrium MBPT on top of DFT, within the local density approximation (LDA), by taking into account the strong SO interaction (see Methods). The quasi-particle band structure has a direct gap at K where the VB and CB SO splittings are $\Delta_v$=455meV and $\Delta_c$=26meV (Fig.\ref{Fig:theory}a). The dynamics as a function of the pump-probe delay $\tau$ is obtained by solving the Kadanoff-Baym equation for the one-body density matrix, $\rho$:
\begin{equation}
\partial_{t}\rho(\tau)=\partial_{t}\rho(\tau)_{coh}+\partial_{t}\rho(\tau)_{coll}
\label{KBE}
\end{equation}
The coherent term, $\partial_{t}\rho(\tau)_{coh}$, includes the electron-hole interaction experienced by the photoexcited electron-hole pairs and describes the interaction with the pump laser, while the dissipative term, $\partial_{t}\rho(\tau)_{coll}$, accounts for all scattering processes. The non-equilibrium population, $f_{i}(\tau)$, is defined as the diagonal part of the single-particle density matrix $f_{i}(\tau)=\rho_{ii}(\tau)$, with $i$ the generic index representing the electronic band and momentum. By solving Eq.\ref{KBE}, we have access to the carrier distribution for all momenta and bands at different $\tau$. In our simulation, the carriers are selectively photoexcited in the K valley by a circularly polarized pump pulse centered around the A exciton. The inset of Fig.\ref{Fig:theory}b is the computed photoexcited CB electron distribution for each Brillouin Zone point for $\tau=0$ (i.e. the maximum of the pump pulse temporal profile). This confirms the valley selectivity by circular polarization. The temporal evolution of the system out of equilibrium is governed by the collision term in Eq.\ref{KBE}, including electron-electron and electron-phonon scattering. 
Electron-electron interaction is expected to increase quadratically with the number of photoinduced carriers\cite{Shah1999}, but we have not measured any appreciable variation of the intravalley dynamics for different excitation fluence (see Fig.\ref{Fig:Fluencedep}). For this reason, we consider a fair approximation to include in $\partial_{t}\rho(\tau)_{coll}$ only the electron-phonon scattering channel. Our assumption is also supported by recent results on 1L-WSe${_2}$, where the intervalley scattering process is well described in terms of electron-phonon scattering\cite{MolinaSanchez2017}.

In the simulation, we focus only on the temporal evolution of the occupation $f_{i}(\tau)$ for the 4 CB states with opposite spin orientation at K/K' $(K,\uparrow)$ $(K,\downarrow)$, $(K',\uparrow)$ and $(K',\downarrow)$ (Fig.\ref{Fig:IntervalleyDynamics}a). The population is first injected in $(K,\uparrow)$ and then scattered towards the other 3 states by phonon mediated processes. The dynamics of $f_{(K,\downarrow)}(\tau)$ (blue trace in Fig.\ref{Fig:theory}b) is governed by phonon-mediated spin-flip intravalley scattering processes. The timescale of the simulated build-up dynamics is close to the experimental one, confirming an efficient scattering rate between upper and lower CB at K. The corresponding $\tau_{rise}$ of $f_{(K,\downarrow)}(\tau)$ decreases at higher temperature as a consequence of the larger phonon population, widening the range of scattering channels available (Fig.\ref{Fig:theory}c,d). The $f_{(K',\downarrow)}(\tau)$/$f_{(K',\uparrow)}(\tau)$ dynamics are related to phonon mediated intervalley scattering processes for photoexcited electrons in CB (red and blue light dashed curves). In this case, the larger electron-phonon matrix elements make the spin-conserving $(K,\uparrow)\rightarrow(K',\uparrow)$ transition faster, even though it involves large momentum phonons. Nevertheless, the scattering towards $(K,\downarrow)$ has a similar fast $\tau_{rise}$, since $(K,\downarrow)$ is the lowest energy state where all carriers are accumulated.

The simulations clearly demonstrate that the inter and intravalley carrier scattering are driven by phonon mediated processes. Nevertheless it is well known\cite{Kaasbjerg2012} that $K\rightarrow K'$ transitions of carriers located {\em exactly} at the bands minimum/maximum are very slow. This evidence is not in contradiction with the present results. Indeed the electron-phonon matrix elements, evaluated between these states at $K$ and $K'$, which correspond to a complete spin flip, are zero. Therefore the $K\rightarrow K'$ dynamics of such localized carriers is much slower that the dynamics observed experimentally. This apparent discrepancy is reconciled by observing that the carriers are, in reality, distributed around $K$ and $K'$. This distribution is induced by several factors: the finite laser energy bandwidth, the broadening of the states\,(proportional to the density of states) caused by quasi--particle corrections\cite{Molina-Sanchez2016}, the fast intravalley scattering that tends to equilibrate carries in the valley by creating a Fermi-Dirac distribution with a finite temperature.
The net consequence is that the inter and intravalley carrier dynamics, shown in Fig.\ref{Fig:theory}(d), is caused by the scattering of carriers {\em around} the $K$ and $K'$ points. It follows that the strength of the interaction and the speed of the transitions is dictated by the density of carriers accumulated around the band minimum/maximum.
It is important, however, to note that our simulations predict build-up times of $f_{i}(\tau)$ which are slower than the measured $\Delta T/T$. This discrepancy can be motivated by several, concurrent phenomena that are not included in our simulation, like the pump-induced optical gap renormalization or the Dexter mechanism\cite{Berghaeuser2018}. Despite this, the electron-phonon interaction remains the driving mechanism which, in addition, well explains the temperature dependence observed experimentally.

In conclusion, we studied intra- and intervalley relaxation dynamics in 1L-WS$_2$ by two-color helicity-resolved transient absorption spectroscopy. This experimental approach allowed us to separately study and to disentangle these processes. We have shown that intravalley relaxation is fast ($<$1ps), leading to a very efficient depletion of the bright A exciton. The temperature dependence of the intravalley scattering rate is a signature of a phonon-mediated process, which depends on the value of the CB spin splitting. Ab-initio simulations based on non-equilibrium MBPT formalism are in good agreement with experiments, confirming the fast timescale. The valley polarization in 1L-WS$_2$ has a similarly fast relaxation time, suggesting a strong interaction between the two valleys. We stress that, since intravalley spin flip and intervalley scattering processes are effective on the same timescale, the former relaxation mechanism cannot be disregarded in explaining the fast valley depolarization of A excitons in 1L-WS$_2$. Since the intravalley scattering process determines the formation of the dark intravalley exciton, our results strongly suggest that the bright to dark intravalley exciton transition occurs on a sub-ps timescale. Dark excitons reduce the light emission efficiency of 1L-TMDs\cite{Malic2018}, therefore, understanding the formation process of dark excitons is of key importance for the design of 1L-TMDs based optoelectronic devices such as light emitting diodes\cite{Palacios2016} and polarized photon emitters\cite{Lodahl2017}.\\
\\

The authors thank Andreas Knorr, Mikhail Glazov, Alexey Chernikov and Ivan Bernal-Villamil for the useful discussions. This work is supported by the National Research Fund, Luxembourg (Projects C14/MS/773152/ FAST-2DMAT and INTER/ANR/13/20/NANOTMD), the Juan de la Cierva Program, the EU Graphene Flagship, ERC grant Hetero2D, EPSRC grants EP/L016087/1, EP/K01711X/1, EP/K017144/1, the Early Postdoc Mobility program of the Swiss National Science Foundation (P2BSP2 168747), the EU project MaX Materials design at the eXascale H2020-EINFRA-2015-1, grant agreement no. 676598, and Nanoscience Foundries and Fine Analysis - Europe H2020-INFRAIA-2014-2015, grant agreement no. 654360.

\section{Methods}

\begin{figure}

\centerline{\includegraphics[width=85mm]{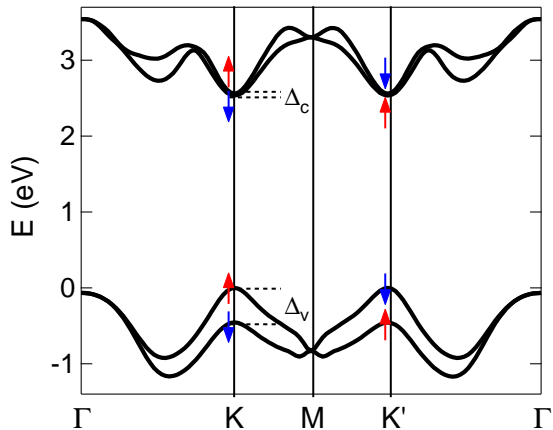}}

\caption{Band structure of 1L-WS$_2$. We have marked the direction $z$-component of the spin at the CB and VB edges at K and K', together with the SO splitting $\Delta_c$ and $\Delta_v$.}

\label{bands}

\end{figure}

\begin{figure}

\centerline{\includegraphics[width=85mm]{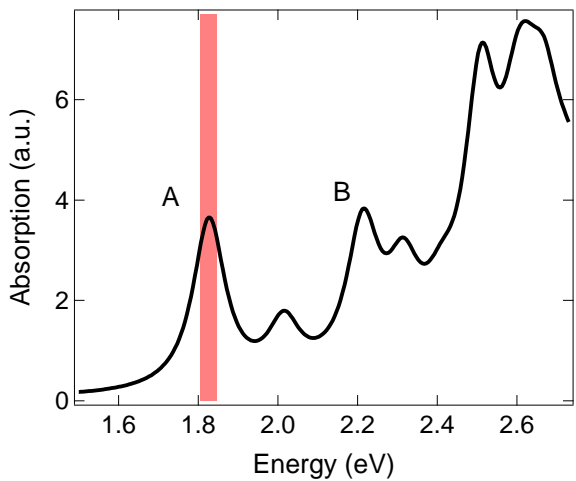}}

\caption{Absorption of 1L-WS$_2$ obtained with the Bethe-Salpeter equation. The A and B excitons are indicated, and the theoretical pump energy is shown with a red band.}

\label{bse}

\end{figure}

\begin{figure}

\centerline{\includegraphics[width=85mm]{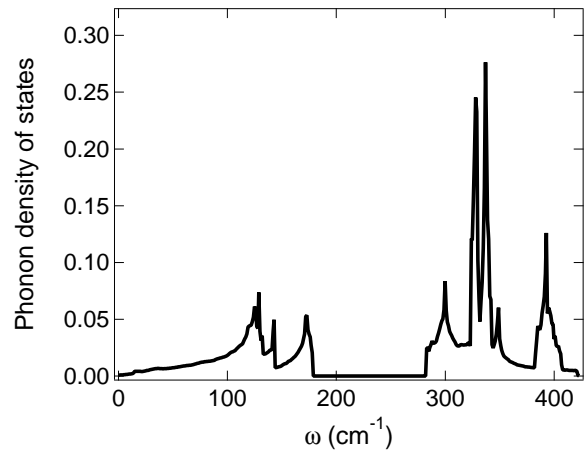}}

\caption{Phonon density of states of 1L-WS$_2$. The frequency gap between low and high phonon modes is 110cm$^{-1}$. At low frequencies the phonon density of states is characterized by two peaks, while at high energy is dominated by a peak centered at$\sim$340cm$^{-1}$, which is associated to the E' mode at $\Gamma$.}

\label{phWS2}

\end{figure}

\subsection{First principles simulations.}

We use the Quantum-Espresso suite\cite{Giannozzi2009} to compute the equilibrium properties of 1L-WS$_2$. The eigenvalues and spinorial wave-functions are calculated with 110Ry cutoff and a $15\times 15\times 1$ $\bf{k}$-point mesh for Brillouin zone sampling. We consider the last two VBs and the first two CBs to describe the carrier dynamics. Electron-phonon matrix elements are computed within Density Functional Perturbation Theory. The real-time simulation is performed with the Yambo code\cite{Marini2009}. The density matrix is propagated in time by solving its equation of motion based on the Kadanoff-Baym equation, within the Generalized Kadanoff Baym ansatz\cite{Marini2013,Sangalli2015a}. The equation of motion for the density matrix is projected on the 4 bands $\rho_{nm\mathbf{k}}$. The real-time simulation adopts a coarse $24\times 24\times 1$ and a denser $61\times 61 \times 1$ $\bf{k}$-grid. The simulation is performed on the denser grid with matrix-elements and density matrix interpolated (by using a nearest neighbor technique) from the coarse grid. The experimental pump pulse is simulated by using a narrow band pump centered at the computed A-exciton position (intensity$\sim10^4$kW/m$^2$, full width at half maximum$\sim$100fs, energy$\sim$1.85eV) The screened-exchange SEX approximation, together with a quasi-particle shift$\sim$1.039eV guarantee that the pump is absorbed at the excitonic peak\cite{Attaccalite2011}. We monitor the diagonal elements of the density matrix, describing the occupations on the band structure, ${f_{n\mathbf{k}}=\rho_{nm\mathbf{k}}}$.
The electronic structure and the optical absorption obtained with the Bethe-Salpeter equation are shown in Figs.\ref{bands} and \ref{bse}. We use DFT in the local density approximation (LDA) as implemented in Quantum Espresso\cite{Attaccalite2011}, with full relativistic pseudopotentials. The optical spectra include the GW correction in the quasi-particle energies and SO interaction using full spinor wave functions. We use a $\bf{k}$-grid of $24\times 24\times 1$ to calculate the Bethe-Salpeter kernel. The position of the $A$ exciton determines the theoretical pump energy to mimic the experiments, where the pump is set to be resonant with A.
The phonon spectrum of 1L-WS$_2$ is calculated using Density Functional Perturbation Theory (DFPT) as implemented in Quantum Espresso\cite{Molina-Sanchez2011}. The optimized lattice parameters are in Table 1 of Ref.\citenum{Molina-Sanchez2011}. The phonon frequencies for different wave-vectors are calculated by solving the secular equation for lattice vibrations, where the coefficients of the dynamical matrix are calculated within DFPT. The calculated phonon density of states of 1L-WS$_2$ is displayed in Fig.\ref{phWS2}.

\end{document}